\documentclass[10pt,conference]{IEEEtran}
\IEEEoverridecommandlockouts
\usepackage{cite}
\usepackage{amsmath,amssymb,amsfonts}
\usepackage{algorithmic}
\usepackage{graphicx}
\usepackage{textcomp}
\usepackage{xcolor}
\usepackage{balance}

\def\BibTeX{{\rm B\kern-.05em{\sc i\kern-.025em b}\kern-.08em
    T\kern-.1667em\lower.7ex\hbox{E}\kern-.125emX}}
    
\begin{document}
\bstctlcite{IEEE-NoDash:BSTcontrol} % for avoiding '_______' in place of authors' names, must have a corresponding entry in bib files, too.

\title{Insights into Dependency Maintenance Trends \\ in the Maven Ecosystem}

\author{\IEEEauthorblockN{
Barisha Chowdhury$^1$  ~~~~~~~~ Md Fazle Rabbi$^2$ ~~~~~~~~ S. M. Mahedy Hasan$^1$ ~~~~~~~~ Minhaz F. Zibran$^2$}
\IEEEauthorblockA{
\textit{$^1$Department of Computer Science, Rajshahi University of Engineering \& Technology, Rajshahi, Bangladesh} \\
\textit{$^2$Department of Computer Science, Idaho State University, Pocatello, ID, United States} \\
%\textit{Idaho State University}\\
%Pocatello, United States \\
nitub81@gmail.com, mdfazlerabbi@isu.edu, mahedy@cse.ruet.ac.bd, zibran@isu.edu}
}

% \and
% \and
% \and
% \and
% \and

\maketitle

\begin{abstract}
% Dependencies are the backbone of modern software development, allowing developers to leverage existing libraries and tools to accelerate project timelines and enhance functionality. However, dependency management is fraught with challenges, particularly in large ecosystems like Maven Central, where millions of libraries and releases coexist.

%The expansion of software ecosystems, particularly through platforms like Maven Central, has resulted in a vast amount of dependency data, offering valuable insights into the structure and dynamics of software projects. 
As modern software development increasingly relies on reusable libraries and components, managing dependencies has become critical for ensuring software stability and security. However, challenges such as outdated dependencies, missed releases, and the complexity of interdependent libraries can significantly impact project maintenance.
In this paper, we present a quantitative analysis of the Neo4j dataset using the Goblin framework to uncover patterns of freshness in projects with different numbers of dependencies. Our analysis reveals that releases with fewer dependencies have a higher number of missed releases. Additionally, our study shows that the dependencies in the latest releases have positive freshness scores, indicating better software management efficacy. These results can encourage better management practices and contribute to the overall health of software ecosystems.

% \rabbi{gave some paper title suggestions after begin{document}. choose anyone of those}
\end{abstract}

\begin{IEEEkeywords}
Dependency management, freshness patterns, Maven, software stability, missed releases
\end{IEEEkeywords}

\section{Introduction}

The growing complexity of software systems has led to more use of external dependencies. These dependencies help developers build software quickly, reduce redundancy, and enhance functionality. They also promote standardization and improve interoperability across different software systems~\cite{sojer2010code}~\cite{holmes2013systematizing}. Repositories such as Maven Central play an important role in the software development ecosystem by hosting millions of libraries and their releases~\cite{raemaekers2013maven}. Developers can access these libraries to speed up innovation and reduce development efforts. The variety of libraries available allows developers to address domain-specific challenges efficiently~\cite{benelallam2019maven}. 

However, using so many dependencies creates problems, such as outdated libraries, missed updates, and compatibility issues between evolving software components. Keeping track of these dependencies and managing them can be difficult. A key but often overlooked factor in dependency management is \textit{freshness}~\cite{cox2015measuring}. Freshness measures how current a release is by looking at how many newer releases are available and how much time has passed since the latest release. This metric is essential for managing dependencies effectively.
% Another important metric is \textit{popularity\_1\_year}~\cite{jaime2024goblin}, which measures how many projects have depended on a release over the past year. 
% Freshness shows how recent a release is, while popularity indicates how widely used it is over time. Both of these metrics are essential for managing dependencies effectively.

Despite the importance of these metrics, there is a lack of studies that explore how freshness varies across different types of dependencies. This gap leaves project maintainers without enough insights into the health of their dependencies, which makes it harder for them to make informed decisions about updates and version management. 

In our study, we investigate how the dependency structures in the Maven ecosystem affect the maintenance and release management of software projects. Specifically, we want to understand if projects with a larger number of dependencies face more challenges in keeping their dependencies up to date and experiencing longer periods of outdated dependencies. We also want to explore how much of the dependencies in the latest releases are outdated, which can highlight the difficulties of maintaining up-to-date software in complex dependency environments. Our main goal is to uncover freshness patterns across a few different areas and address the following research questions (RQs):

% Our study investigates how the dependency structures in Maven Central affect the maintenance and release management of software projects. Specifically, we seek to understand whether projects with a larger number of dependencies encounter more difficulties in keeping their dependencies up to date and experience longer periods of outdated dependencies. We also aim to explore the extent to which dependencies in the most recent releases are outdated, shedding light on the challenges of maintaining up-to-date software in complex dependency ecosystems. The main goal is to uncover freshness patterns across a few dimensions while addressing the following research questions (RQs):

%\vspace{0.1cm}
\textbf{RQ1:} Do projects with a large number of dependencies tend to have a higher ``outdated time" or missed releases compared to those with fewer dependencies?

-- The findings from this question can provide valuable insights into how the number of dependencies affects project maintenance. By understanding the relationship between dependency counts and outdated dependencies, developers can optimize update processes, reduce delays, and ensure that dependencies remain up-to-date. This can lead to more efficient and stable project management.

% -- This finding can help the developer community by identifying the impact of dependency counts on project maintenance. It could provide insights into how to streamline updates, making projects more resilient to delays and ensuring dependencies stay current.

%\vspace{0.1cm}
\textbf{RQ2:} To what extent are the dependencies in the latest software releases outdated?

-- If dependencies in recent releases are found to be outdated, it may indicate gaps in update practices or challenges in maintaining compatibility with newer versions. This could have implications for software stability and security. Understanding these trends can help developers prioritize critical updates, improve dependency management strategies, and make more informed decisions when choosing or maintaining dependencies.

% -- Outdated dependencies in recent releases may indicate inefficiencies in update practices or challenges in maintaining compatibility with newer versions, posing risks to software stability and security. Identifying trends in outdatedness in the latest releases can help developers prioritize updates and make more informed decisions about adopting or maintaining specific dependencies.
%\vspace{0.1cm}

To address the aforementioned RQs, we use the Goblin Miner tool~\cite{jaime2024goblin} and the Maven Central Neo4j dataset~\cite{jaime2024neo4j}. By using this dataset and tool, we aim to highlight the challenges and trends in managing dependencies within the Maven Central ecosystem. A comprehensive replication package~\cite{replicationPackage}, including datasets, analysis scripts, and documentation, has been provided to ensure the reproducibility of this study’s findings.

% To facilitate the dynamic analysis of dependency structures, the Goblin Miner tool~\cite{jaime2024goblin} and  Maven Central Neo4j dataset are used~\cite{jaime2024neo4j}. By leveraging this rich dataset and toolset, this paper aims to shed light on the challenges and trends in managing dependencies within the Maven Central ecosystem.

%The rest of the paper is organized as follows. In Section~\ref{sec:data}, we introduce the dataset used in this study. In Section~\ref{sec:result}, we present our analysis and findings. Section~\ref{sec:threats} discusses the possible limitations of this work. Related work is covered in Section~\ref{sec:related}. Finally, we conclude the paper in Section~\ref{sec:conclusion}.

\vspace{-0.2cm}
\section{Dataset}\label{sec:data}

For this study, we use the publicly available Neo4j Maven Central dependency graph dataset~\cite{jaime2024neo4j}. We work with the ``with\_metrics\_goblin\_maven\_30\_08\_24.dump," dated August 30, 2024. The database contains over 15 million nodes, including 658,078 libraries (artifacts) and 14,459,139 releases. It also includes 134 million edges, with 119,660,406 dependencies and 14,459,139 versioning relationships. The dataset adopts a dependency graph model with two main types of nodes: Artifacts (libraries) and Releases (specific versions of artifacts). Edges in the graph represent release-to-artifact (R→A) and artifact-to-release (A→R) relationships, showing dependency and versioning structures. The dataset also includes additional metadata, such as version ranges, release timestamps, and scopes (e.g., compile, test), which allow for detailed analysis.

The graph structure allows us to compute various metrics like freshness, release rhythm, and vulnerability exposure (CVE data) on-demand using the Goblin framework’s Weaver component. This flexibility makes the dataset suitable for exploring key questions about dependency freshness, release patterns, and ecosystem health.

To ensure the feasibility of the study, we analyze a subset of the data, focusing on 100,000 libraries and 1,000,000 dependencies. This allows us to examine dependency management practices while maintaining computational efficiency. This dataset forms the basis for investigating trends and challenges in software dependency management within large ecosystems.

%The graph structure supports various metrics, including freshness, release rhythm, and vulnerability exposure (CVE data), which can be computed on-demand through the Goblin framework’s Weaver component. This flexibility makes the dataset particularly suitable for investigating key questions related to dependency freshness, release regularity, and ecosystem health.

%To ensure feasibility for the study, a subset of 100,000 libraries and 2,000,000 dependencies was analyzed, enabling focused exploration of dependency management practices while maintaining computational efficiency. This dataset provides the foundation for investigating trends and challenges in software dependency management within large-scale ecosystems.

% The dataset's granularity and scale enable in-depth analyses of dependencies, outdatedness, and versioning dynamics across the ecosystem. Some predefined metrics are included: Common Vulnerabilities and Exposures (CVEs), Freshness, Popularity and Release Rhythm. These metrics are incorporated into the database as AddedValue edges, allowing for enriched analyses. 

\section{Analysis and Results}\label{sec:result}
\subsection{Dependency Counts and Freshness}
\subsubsection{Methodology}

We use the \textit{dependency} relationship between releases and artifacts to determine the number of dependencies a release possesses. To assess freshness, we extract different freshness scores using the \textit{AddedValue} edge. The dataset contains 119,660,406 dependencies, but for computational convenience, we analyze a subset of 1,000,000 dependencies ($\approx $ 0.84\% of the total). This subset of dependencies comes from 107,916 releases, with the minimum number of dependencies for a single release being 1 and the maximum being 417. By selecting this smaller subset, we balance computational efficiency with the need to capture a broad range of dependency management practices. Although the subset represents only 0.84\% of the total dependencies, it still includes data from over 107,000 releases, which is large enough to reflect diverse patterns and trends within the dataset.  Dependencies in software ecosystems often follow predictable patterns~\cite{Decan2017An}, making this subset large enough for meaningful analysis.

%Taking advantage of the dependency relationship between release and artifact (R→A), we determine how many dependencies one release possesses. Then, by AddedValue edge, different freshness scores were obtained. As stated earlier, there are 119,660,406 dependencies altogether. Among these, we consider 10,00,000 dependencies for our convenience. Since dependencies in software ecosystems often exhibit predictable patterns~\cite{Decan2017An} and analyzing a smaller dataset allows for a deeper investigation into specific aspects of dependency management, 10,00,000 dependencies seemed large enough to capture diverse patterns and trends present in the dataset. The said number of dependencies represents 1,07,916 releases. The minimum and maximum dependencies for a single release are 1 and 417, respectively. 

\begin{figure}[htbp]
    \centering
    \includegraphics[width=0.48\textwidth]{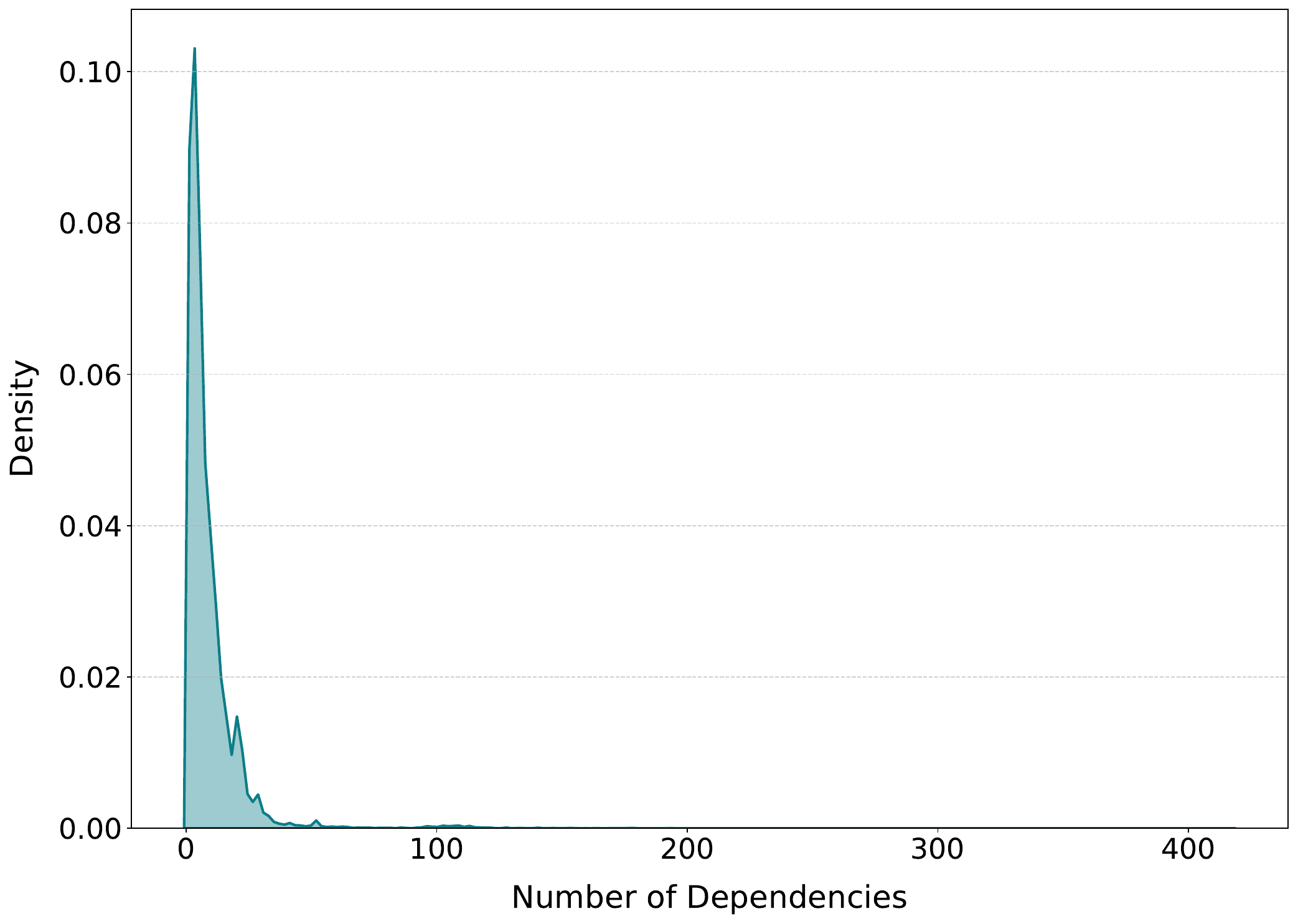}
    \vspace{-0.3cm}
    \caption{Distribution of Dependencies}
    \label{fig:dis_dep}
    \vspace{-0.1cm}
\end{figure}

\begin{figure}[htbp]
    \centering
\includegraphics[width=0.48\textwidth]{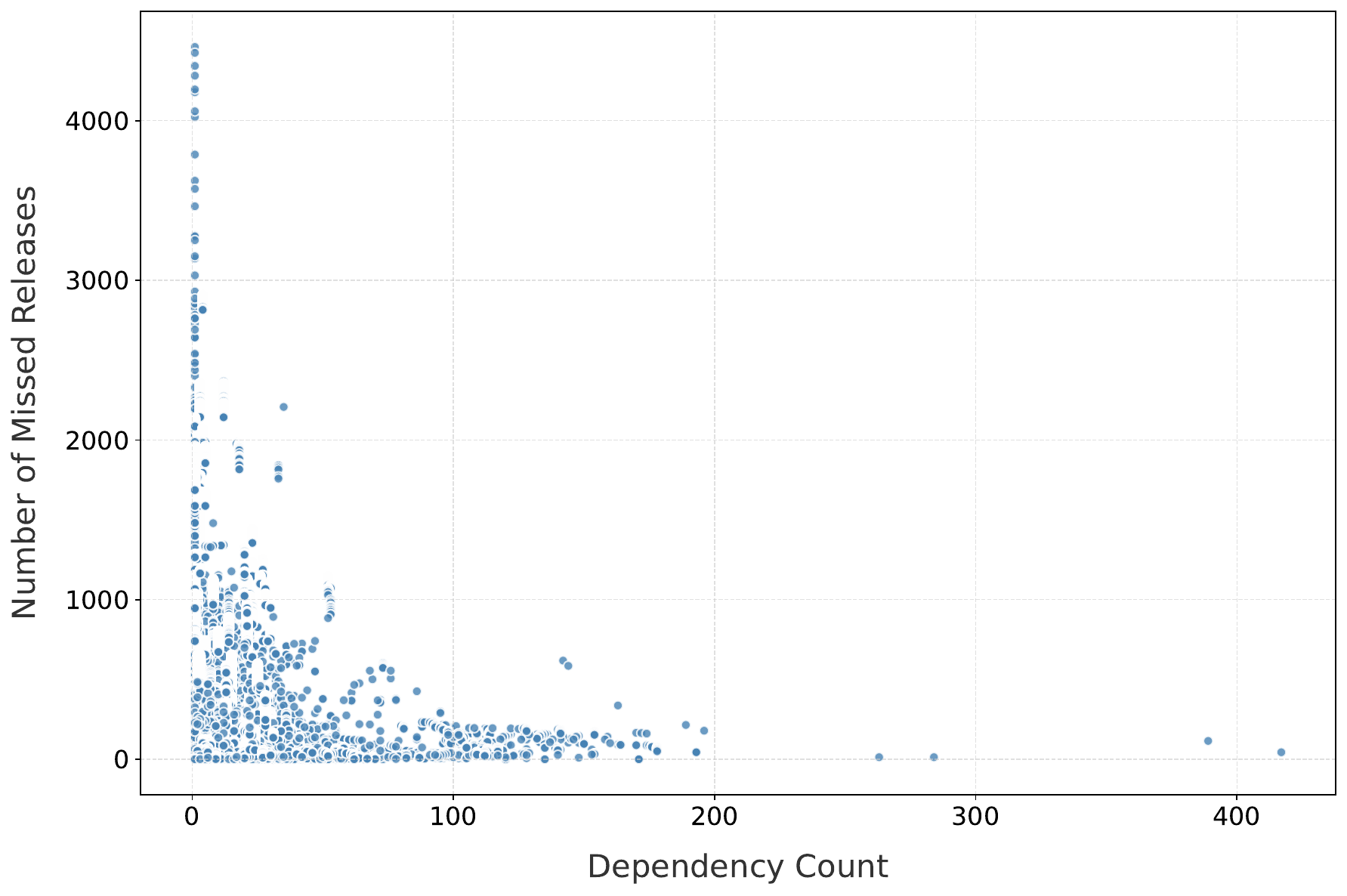}
    \vspace{-0.3cm}
    \caption{Dependency Count Vs Number of Missed Releases}
    \label{fig:missed_rel}
    \vspace{-0.1cm}
\end{figure}

\subsubsection{Findings}

Figure~\ref{fig:dis_dep} represents the Kernel Density Estimation (KDE)~\cite{chen2017tutorial} plot for the distribution of dependencies. We observe that most of the data is concentrated around the lower values on the x-axis, near 0 dependencies. This indicates that most releases have very few dependencies. The long tail on the right shows that a small number of releases have very high numbers of dependencies, but these are outliers and occur less frequently. The peak density is slightly above 0.10, representing around 10\% of the data.

Figure~\ref{fig:missed_rel} shows the relationship between dependency count and the number of missed releases across projects. We observe that most projects have fewer than 50 dependencies. These projects have a wide range of missed releases, from zero to over 4000. We notice that projects with high numbers of missed releases, such as those exceeding 1000, often have low to moderate dependency counts. This suggests that missed releases are not always caused by a high number of dependencies.

% (previous) We find that projects with fewer dependencies, which make up nearly 80\% of the dataset, have a lot of variation in missed releases. These projects often miss around 1500 releases on average. Smaller projects often lack the resources or maintenance practices needed to update dependencies on time. Projects with 50 to 200 dependencies tend to show more consistent updates, averaging around 500 missed releases. Projects with over 200 dependencies typically miss fewer than 100 releases. This shows that, despite their complexity, projects with more dependencies manage updates more effectively.

%We categorize dependency counts into three groups: low (fewer than 50 dependencies), moderate (50 to 200 dependencies), and high (more than 200 dependencies). Projects with fewer dependencies, which make up nearly 80\% of the dataset, have a lot of variation in missed releases, often averaging around 1500 missed releases. Smaller projects often lack the resources or maintenance practices needed to update dependencies on time. Projects with moderate dependencies tend to show more consistent updates, averaging around 500 missed releases. High-dependency projects typically miss fewer than 100 releases, indicating that, despite their complexity, these projects manage updates more effectively.

We find that projects with fewer (less than 50) dependencies, which make up nearly 80\% of the dataset, have a lot of variation in missed releases. These projects often miss around 1500 releases on average. Smaller projects often lack the resources or maintenance practices needed to update dependencies on time. Projects with moderate (50 to 200) dependencies tend to show more consistent updates, averaging around 500 missed releases. High-dependency (over 200) projects typically miss fewer than 100 releases. This shows that, despite their complexity, projects with more dependencies manage updates more effectively.

% From Figure~\ref{fig:missed_rel}, it is discernible that the majority of projects are concentrated at the lower end of the dependency count axis (fewer than 50 dependencies). These projects exhibit a wide range of missed releases, spanning from zero to over 4000. Projects with very high numbers of missed releases (e.g., exceeding 1000) are mostly associated with low to moderate dependency counts, suggesting that missed releases are not necessarily a result of extremely high dependency counts. A few projects with dependency counts exceeding 200 have relatively low numbers of missed releases, indicating potential cases of well-maintained projects or projects with fewer disruptions.

\begin{figure}[htbp]
    \centering
    \includegraphics[width=0.48\textwidth]{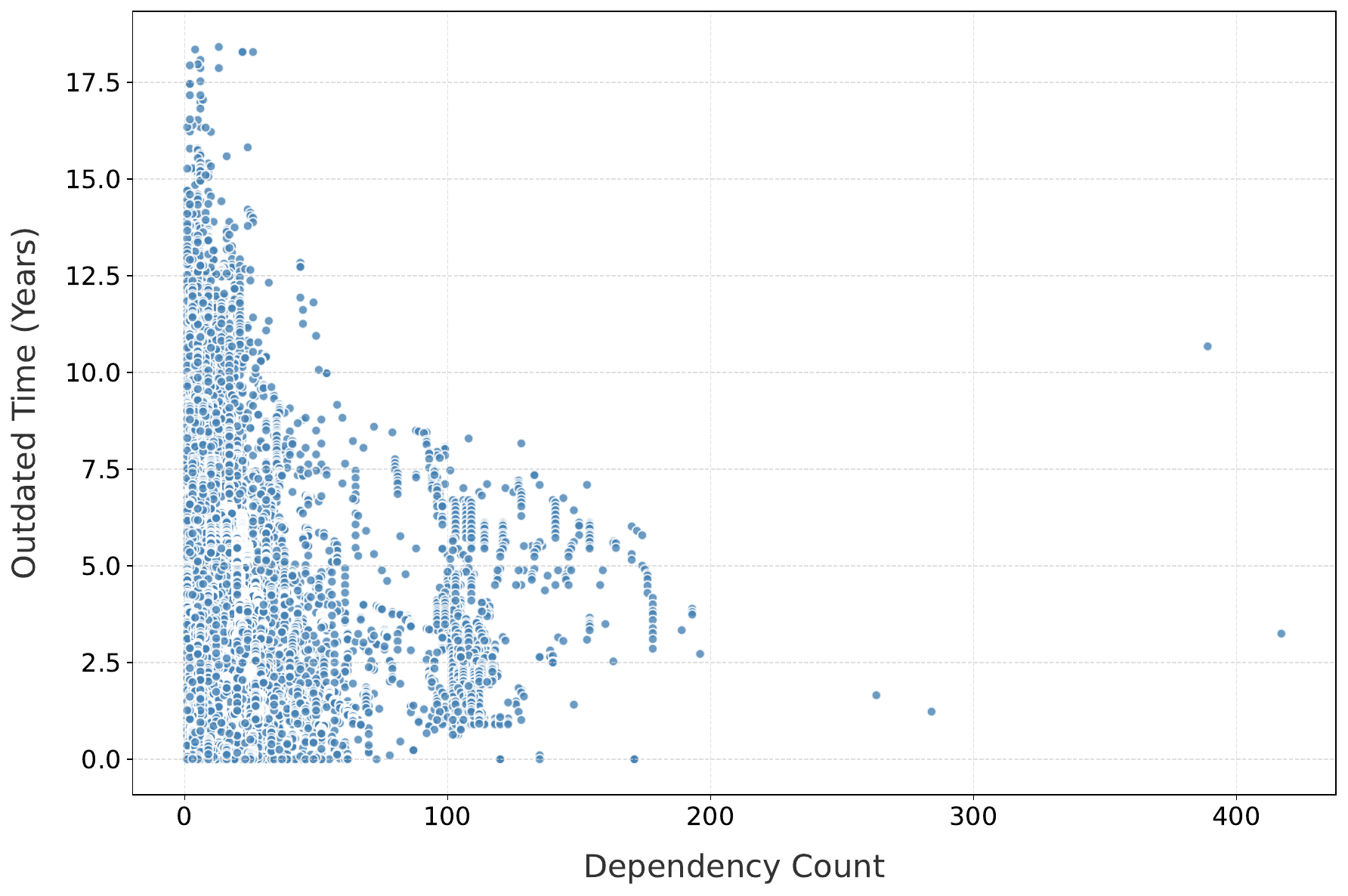}
    \vspace{-0.2cm}
    \caption{Dependency Count Vs Outdated Time (Years)}
    \label{fig:out_time}
    \vspace{-0.1cm}
\end{figure}

% (Previous) Further supporting these findings, Figure~\ref{fig:out_time} shows the outdated times of dependencies for projects with different dependency counts. We find that low-dependency projects have outdated times ranging up to 17.5 years, with an average of about 6 years. Projects with moderate dependency counts show better control, with an average outdated time of about 2.5 years. High-dependency projects have the lowest outdated times, averaging less than 1 year, with most dependencies staying current. These projects benefit from regular release cycles, automated dependency tracking, and proactive updates. 

Further supporting these findings, Figure~\ref{fig:out_time} shows the outdated times of dependencies for projects with different dependency counts. We find that low-dependency projects have outdated times ranging up to 17.5 years, with an average of about 6 years. Projects with moderate dependency counts show better control, with an average outdated time of about 2.5 years. High-dependency projects have the lowest outdated times, with most staying under 2 years, having some exceptions. These projects benefit from regular release cycles, automated dependency tracking, and proactive updates. 

% Figure~\ref{fig:out_time} suggests the same as Figure~\ref{fig:missed_rel} and that is releases with fewer dependencies exhibit a wide range of outdated times, extending up to 17.5 years. This conveys that even small projects can have significantly outdated dependencies.

The results show an inverse relationship between dependency counts and maintenance challenges. Smaller projects tend to struggle with more missed releases and longer outdated times. Larger projects, however, benefit from structured and proactive dependency management.

% The findings revealed that projects with fewer dependencies tend to have more frequent missed releases and the majority of projects with high dependency counts (greater than 200) demonstrated minimal missed releases. This might be attributed to the fact that smaller projects often lack the resources, attention, or rigorous maintenance practices required to manage timely updates. Additionally, fewer dependencies may indicate that such projects or releases are less actively developed or maintained, contributing to delayed or missed releases. On the other hand, high-dependency projects are often part of larger ecosystems, where regular updates are crucial to maintaining the overall reliability of software. These projects likely adopt automated tools, such as dependency management frameworks, to monitor and address outdated or missed releases, thereby reducing the frequency of missed releases. 

\subsection{Dependency of Latest Releases and Freshness}
\subsubsection{Methodology}
To identify the latest releases for each artifact, we use the \textit{relationship\_AR} edge, which connects artifacts to their releases. We analyze the dependencies of these latest releases and their freshness using the dependency edge along with the \textit{AddedValue} edge. Out of the 658,078 artifacts in the dataset, we select a subset of 100,000 libraries ($\approx$ 15.2\%) to ensure computational feasibility while maintaining statistical validity, as recommended in prior studies~\cite{Leskovec2006Sampling}. The selected libraries contain 742,492 dependencies. However, freshness data is unavailable for 32,066 dependencies, about 4.3\% of the total. As a result, we proceed with the remaining 710,426 dependencies (95.7\%).

% To get the latest releases of every artifact, we work with relationship\_AR edge between artifact and release (A→R). For the dependencies of the latest releases and the freshness of those dependencies, we utilize the dependency and the AddedValue edge. Out of the 658,078 artifacts available in the dataset, 100,000 libraries were chosen to ensure computational feasibility and maintain a balance between resource utilization and data comprehensiveness. This sample size aligns with guidelines for statistical validity in large-scale graph analyses~\cite{Leskovec2006Sampling}. The latest releases collectively include a total of 742,492 dependencies. Since freshness data was unavailable for 32,066 of them, we proceeded to examine the freshness score of 710,426 releases. 

\subsubsection{Findings}
\begin{figure}[htbp]
    \centering
    \includegraphics[width=0.48\textwidth]{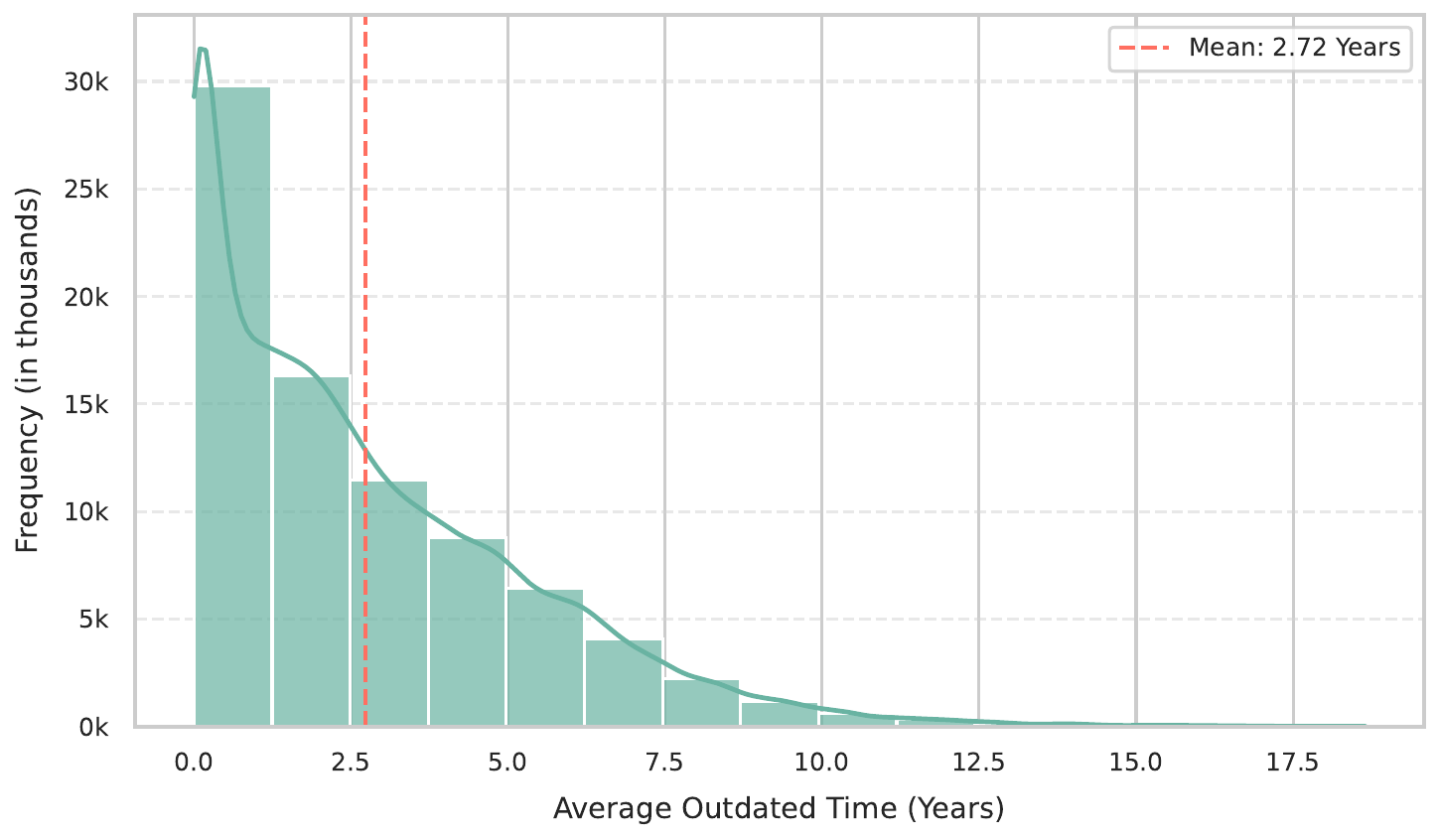}
    \vspace{-0.2cm}
    \caption{Distribution of Outdated Time (Years) Across Dependencies}
    \label{fig:out_time2}
\end{figure}

Figure~\ref{fig:out_time2} shows the distribution of outdated times across the dependencies of the latest releases. We see a sharp concentration near zero, which indicates that most dependencies are up-to-date. However, a small subset of dependencies has significantly higher outdated times, with some exceeding multiple years. The mean outdated time for dependencies is 2.5 years. This means that while many dependencies are current, some experience delays in updates.

% Figure~\ref{fig:out_time2} suggests that dependencies of the latest releases tend to stay fairly up-to-date in most cases. There are a small number of cases with significantly higher outdated times. Since the data set pertains to a system that requires timely updates, a high concentration near zero is a positive sign of efficiency. 

\begin{figure}[htbp]
    \centering
    \includegraphics[width=0.48\textwidth]{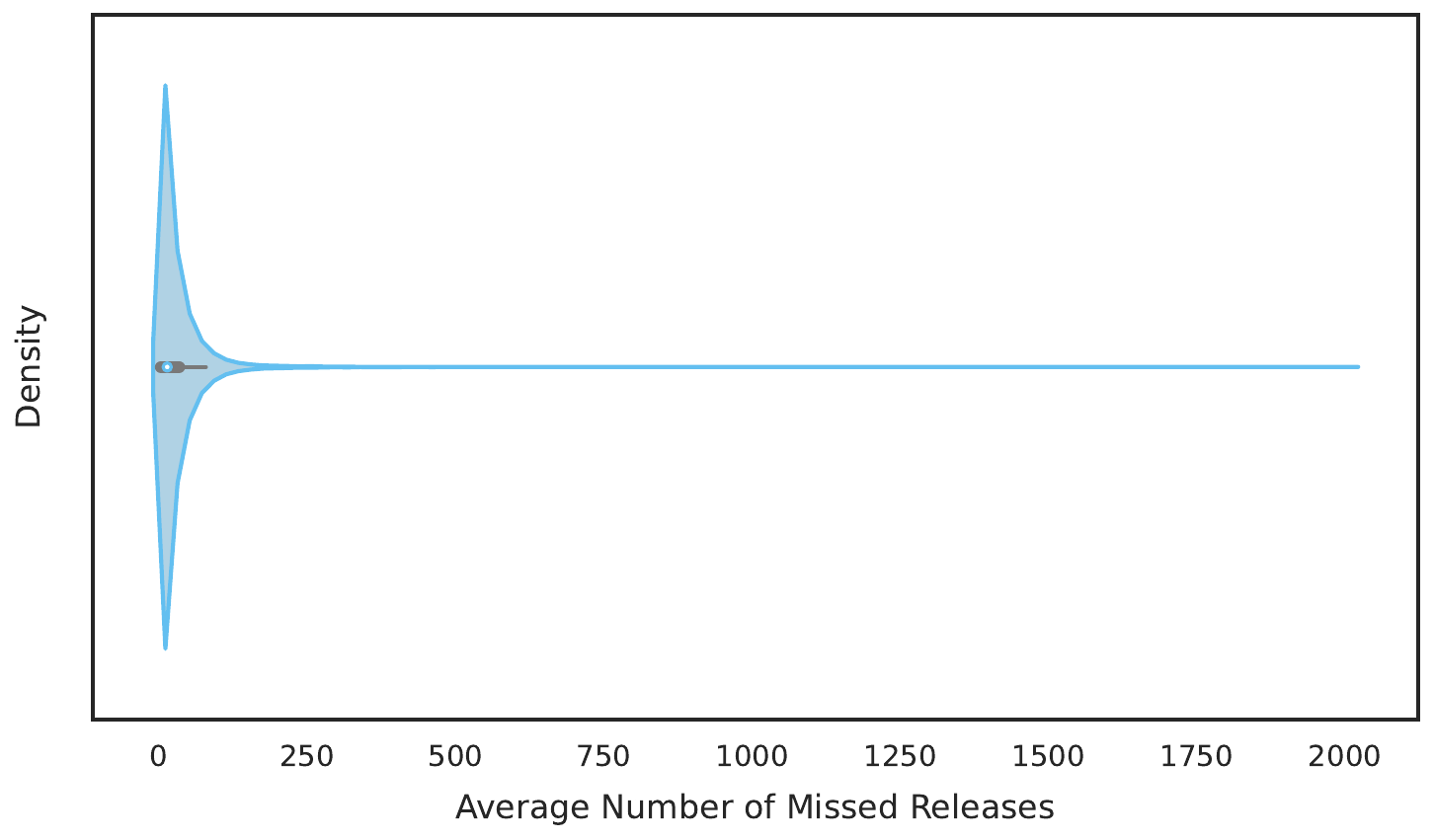}
    \vspace{-0.2cm}
    \caption{Distribution of Missed Releases Across Dependencies}
    \label{fig:missed_rel2}
    \vspace{-0.1cm}
\end{figure}

Figure~\ref{fig:missed_rel2} shows a similar distribution for missed releases across dependencies. The density is concentrated near zero, which suggests that most dependencies have few or no missed releases. However, there are some dependencies with a large number of missed releases, but these are relatively rare.

% Figure~\ref{fig:missed_rel2} expresses the identical outcome as Figure~\ref{fig:out_time2}. The density is sharply concentrated near zero, indicating that the majority of the dependencies of the latest releases have few missed releases. This suggests that missed releases are relatively uncommon for most dependencies. 

The results suggest that dependencies in the latest releases are generally well-maintained, with minimal missed releases and short outdated times for most cases. However, some dependencies have long outdated times and numerous missed releases, indicating that updates have been neglected in these cases. Whether a mean outdated time of 2.5 years is problematic depends on the project’s needs and how it manages dependencies. In fast-evolving domains like web frameworks and cloud applications, updates are needed every few months to keep up with changing technologies and security requirements~\cite{Bhattacharya2010Dynamic}. In these cases, a 2.5-year delay could lead to significant technical debt. For more stable or legacy systems, however, this delay might be acceptable, as older dependencies can still function effectively~\cite{Richmond2006An}~\cite{Machida2017Lifetime}.

\section{Threats to Validity}\label{sec:threats}
Our methodology provides a quantitative analysis of dependency management trends based on a large-scale dataset. However, we must acknowledge several limitations to help contextualize our findings and their applicability.

We rely on a subset of the Maven Central dataset, consisting of 100,000 libraries and 1,000,000 dependencies. This raises concerns about how well our results apply to the broader dataset. While we chose this subset for computational feasibility, it may not capture patterns found in the larger dataset with millions of artifacts and releases. This limitation could overlook trends in less commonly used libraries or niche areas within the ecosystem. Additionally, we focus only on Maven Central, which, while widely used, is just one software repository. Other ecosystems, such as PyPI or npm, may have different dependency structures and practices, which limits the generalizability of our findings.

Another limitation comes from the simplicity of the metrics we use. Metrics like outdatedness and missed releases provide valuable insights but do not capture more complex aspects of dependency management. For instance, indirect dependencies, scope-specific analyses (e.g., test vs. compile dependencies), and varying update policies could offer a deeper understanding of the challenges developers face.

Additionally, our results are based on data-driven analysis and lack validation from case studies or developer interviews. Qualitative approaches could provide richer insights and confirm the observed trends, especially correlations like those between dependency counts and missed releases. Another potential bias arises from how we select libraries for the analysis. We choose libraries based on feasibility criteria, which may skew the sample toward well-maintained or frequently updated projects.
Lastly, we do not account for external factors that may influence dependency management and release practices. Community size, project funding, and developer engagement can all affect how well dependencies are maintained and updated.

\section{Related Works}\label{sec:related}
%Dependencies are fundamental to modern software development but managing dependencies is far from trivial. 

Many studies in the past involved the investigation of bug patterns~\cite{Islam_BugPattern_SAC2020,Islam_BugPattern_ACR2021,Amit_ChromiumBug_2022}, 
vulnerabilities~\cite{Islam_2016_IWSC,Islam_2017_ESEM}, code smells~\cite{Islam_IWSC_2018,Zibran_2013_RefactorSchedule,Zibran_2013_CloneChangeStudy}, code quality~\cite{Duaa_2018_ComplexityReadability,Islam_IWSC_2018}, human aspects~\cite{Rabbi_2023_PhoneSensor,Champa_2023_Female,Islam_2016_IJSI,Islam_2016_SERA,Islam_BugSentiment_SEDE2018} of software development and maintenance as well as comparison of methods/tools~\cite{Islam_2017_DictionaryConstruction,Islam_2018_SentiToolCompare,Ryan_AndroidSec_2021,Daniel_WordPress_2021} for measuring such aspects.

Software dependency management has become a key focus in software engineering, especially in large ecosystems like Maven Central. The development of frameworks like Goblin provides access to extensive dependency graph data. This enables real-time analysis of patterns and trends and broadens the scope of research. We can now explore the effects of dependency management practices, outdated libraries, and missed releases across various projects and domains globally.

Assessing how the number of dependencies affects the project is crucial due to its relevance to software maintenance and the ecosystem. Cataldo et al.~\cite{cataldo2009software} analyzed different types of dependency data from two independent software projects over an eight-year period. The analysis suggested software systems with a large number of dependencies, particularly logical and work dependencies, tend to be more complex and can lead to challenges in managing the development process. One of their research questions also revealed that as the inter-dependencies among tasks increase, the likelihood of defects in the software also rises. Tellnes~\cite{tellnes2013dependencies} showed that the security and availability of a system are largely determined by the surrounding `ecosystem' of dependencies. Prana et al.~\cite{prana2021out} highlights the importance of managing the number of dependencies and performing timely updates.

Several studies have demonstrated the necessity of specific metrics for quantifying dependency freshness and how it evolves. Cox et al.~\cite{cox2015measuring} performed correlation and longitudinal analysis to investigate the relationship between dependency freshness and known security vulnerabilities and to assess the variability of the metric over time, respectively. Kula et al.~\cite{raula_gaikovina_kula__2017} emphasized the importance of keeping dependencies updated, proposing a Software Universe Graph (SUG) to model dependency relationships and provide metrics for assessing update needs. Jafari et al.~\cite{javan2023dependency} investigated how different package characteristics can influence the predicted update strategy and found dependent count to be one of the highest influencing features. The results of Zerouali et al.~\cite{zerouali2018empirical} show the strong presence of technical lag and reluctance caused by the specific use of dependency constraints.

After evaluating the relevant studies, we find that there is room for further contributions in understanding the characteristics of dependencies and the freshness of projects. While existing literature explores dependency management, most studies focus on broad trends such as dependency growth, versioning policies, or security vulnerabilities. This leaves gaps in understanding whether there is a direct connection between the number of dependencies and the freshness of software projects, especially in the Maven ecosystem. Limited attention has been given to evaluating how up-to-date the dependencies in the latest releases are, which we aim to address in our study.

\section{Conclusion}
\label{sec:conclusion}
This study presents an empirical analysis of dependency management in the Maven Central ecosystem, driven by two core research questions: whether projects with more dependencies are more likely to miss releases, and to what extent the dependencies of the latest releases are outdated. We use quantitative insights to analyze 100,000 libraries and over 1,000,000 dependencies. Our findings reveal that projects with fewer dependencies are more likely to miss releases, while projects with more than 200 dependencies tend to have fewer missed releases. We also find that the dependencies in the latest releases are generally up-to-date, indicating proactive management in current software development practices.
With the increasing complexity of software ecosystems, this research provides actionable insights into the challenges of dependency management. Understanding patterns of missed releases and outdated dependencies can inform strategies to improve release reliability and dependency maintenance. However, it is important to acknowledge limitations such as dataset constraints and the scope of library selection, which may not fully capture the broader ecosystem of dependencies. Additionally, we did not include qualitative data, such as developer interviews or case studies, which could provide deeper insights into the challenges and strategies behind dependency management. 

Future work could explore dependency management practices across diverse ecosystems or investigate the role of external factors, such as developer collaboration like direct surveys or interviews with them and release policies, in shaping dependency health. Real-time modeling of dependency release trends could also reveal evolving practices in managing software ecosystems. Ultimately, this research highlights key dependency trends and their implications, contributing to more resilient and efficient software engineering practices.

% \rabbi{add latest papers (2020-2024 year) in the reference section. Currently, there are no such papers}

% \rabbi{Enlarge the whole paper by about half a page so that the reference section starts on the 5th page.}

% \section*{Acknowledgment}

% The preferred spelling of the word ``acknowledgment'' in America is without 
% an ``e'' after the ``g''. Avoid the stilted expression ``one of us (R. B. 
% G.) thanks $\ldots$''. Instead, try ``R. B. G. thanks$\ldots$''. Put sponsor 
% acknowledgements in the unnumbered footnote on the first page.

\balance

\section*{Acknowledgement}
This work is supported in part by the ISU-CAES (Center for Advanced Energy Studies) Seed Grant at the Idaho State University, USA.

\bibliographystyle{IEEEtran}
\bibliography{Reference,Bug,SBOM_JS,sentiment,phishing,gender}
\end{document}